# EFFECT OF QUERY FORMATION ON WEB SEARCH ENGINE RESULTS

Raj Kishor Bisht[1] and Ila Pant Bisht[2]

[1]Department of Computer Science & Applications, Amrapali Institute, Haldwani (Uttarakhand), India
`bishtrk@gmail.com`

[2] Dept. of Economics & Statistics, Govt. of Uttarakhand, Divisional Office, Haldwani, (Uttarakhand) India
`Pant_ila@rediffmail.com`

## ABSTRACT

*Query in a search engine is generally based on natural language. A query can be expressed in more than one way without changing its meaning as it depends on thinking of human being at a particular moment. Aim of the searcher is to get most relevant results immaterial of how the query has been expressed. In the present paper, we have examined the results of search engine for change in coverage and similarity of first few results when a query is entered in two semantically same but in different formats. Searching has been made through Google search engine. Fifteen pairs of queries have been chosen for the study. The t-test has been used for the purpose and the results have been checked on the basis of total documents found, similarity of first five and first ten documents found in the results of a query entered in two different formats. It has been found that the total coverage is same but first few results are significantly different.*

## 1. INTRODUCTION

A web query is a set of words or a single word that a searcher enters into the web search engine to get some information as per his or her requirement. Web search queries entered by web searcher are unstructured and vary from standard query languages. A common searcher enters a query into web search engine according to his or her own way of communication. For example, to know about economy of India, two queries "Economy of India" and "Indian Economy" can be put. Though both the queries are semantically same but syntax of both are different a little bit. As far as key words are taken into consideration, after removing stop words and stemming, both the queries have same content words "India" and "Economy". The searcher expects same results in both of the cases as both the queries are semantically same and also contain same content words. But in general, it is observed that the search engine does not provide same results for a query entered in two different forms, however some documents are common in two results. In this paper, we have studied the effect of query formation on web search engine results in terms of coverage of documents and similarity of first five and first ten documents. We select Google search engine for our experiment due to its popularity. So far many researchers have investigated the behavior of web search results and effect of query formation on them. Some interesting characteristics of web search have been showed [7] by analyzing the queries from the Excite search engine like, the average length of a search query was 2.4 terms, about half of the users entered a single query while a little less than a third of users entered three or more unique queries, close to half of the users examined only the first one or two pages of results (10 results per page), less than 5% of users used advanced search features (e.g., Boolean operators like AND, OR, and NOT) etc. Study shows that librarians may not routinely be teaching queries as a strategy for selecting and using search tools on the Web [1]. Karlgren, Sahlgren and Cöster [5] investigated topical dependencies between query terms by analyzing the distributional character of query terms. Topi and Lucas [8] examined the effects of the search interface and Boolean logic training on user search performance and satisfaction. Topi and Lucas [9] presented a detailed analysis of the structure and components of queries written by experimental participants in a study that manipulated two factors found to affect end-user information retrieval performance: training in Boolean logic and the type of search interface. Vechtomova and Karamuftuoglu [10] demonstrated effective new methods of document ranking based on lexical cohesive relationships between query terms. Eastman and Jansen [2] analyzed the impact of query operators on web search engine results. One can find the detail of information retrieval technology in the book of Manning, Raghavan, and Schutze [6] .

The structure of the paper is as follows: Section 2 describes the research design and methodology. In Section 3, experimental results are given and finally section 4 describes conclusions of the study.

## 2. RESEARCH METHODOLOGY

This section describes the specific research questions and the methodology used for study.

### 2.1. Research Question

The present study investigates the following research questions:

1) Is there any change in coverage (total no. of documents found) of results retrieved by Google search engine in response to semantically same but two different forms of a query?

Here the objective is to check the difference in number of documents retrieved in response to two forms of a query. Google search engine provides the total no. of results found against a query. Since a searcher may search the information in any of the documents, thus it is important to know whether the coverage of two results is same or not. The null and alternative hypotheses are as follows:

Null Hypothesis: There is no difference in the coverage.
Alternative hypothesis: The coverage of two results is significantly different.

2) Whether the first few documents (5 or 10) are same in the two results retrieved by Google search engine in response to semantically same but two different forms of a query?

Study shows that approximately 80% of web searchers never view more than the first 10 documents in the result list [3,4]. Based on this overwhelming evidence of web searcher behaviour, we utilized only the first 5 and 10 documents in the result of each query. We have checked the number of documents common in sample queries. Assuming that the first five and first ten documents are same in two results, population mean can be taken as five and ten respectively. The null and alternative hypotheses are as follows:

Null Hypothesis: First 5 and first 10 documents are same in two results, that is, sample mean is equal to population mean.

Alternative hypothesis: First 5 and first 10 documents are significantly different in two results, that is, the sample mean is significantly different from population mean.

We choose 5% level of significance for inference.

## 2.2. Methodology

For first problem, we shall use paired t-test as it can be assumed that the difference of number of observations distributed normally. Let $D_i$ denotes the difference of two observations of $i^{th}$ pair. Under the null hypothesis $H_0$ that there is no significant difference between the two observations, the paired t-test with n-1 degree of freedom is the test statistics

$$t = \frac{\overline{D}}{S/\sqrt{n}} \qquad (1)$$

where $\overline{D} = \frac{1}{n}\sum_{i=1}^{n} D_i$, $S^2 = \frac{1}{n-1}\sum_{i=1}^{n}(D_i - \overline{D})^2$ and $n$ be the number of observations taken.

For first problem, Google search engine shows the number of documents retrieved in response to a query. Let $x_i$ and $y_i$ be the number of documents retrieved in two forms of $i^{th}$ query. In this case $D_i$ is the difference of $x_i$ and $y_i$.

For second problem, let $\overline{x}$ be the mean of the sample of size $n$, $\mu$ be the population mean, $S^2$ be the unbiased estimate of population variance $\sigma^2$, then to test the null hypothesis that the sample is from the population having mean $\mu$, the student's t- test with $n-1$ degree of freedom, is defined by the statistics

$$t = \frac{\overline{x} - \mu}{S}\sqrt{n-1} \qquad (2)$$

Where $\overline{x} = \frac{1}{n}\sum_{i=1}^{n} x_i$ and $S = \frac{1}{n-1}\sum_{i=1}^{n}(x_i - \overline{x})^2$.

## 2. EXPERIMENTAL RESULTS

Fifteen pairs of queries have been farmed on general basis (see appendix A). The queries have been submitted to the search engine from 10$^{th}$ May 2012 to 19$^{th}$ May 2012. Results of every pair of query have been noted down. For each query, it has been observed that all retrieved documents were not same in two forms and also the order of common retrieved documents were different in two results. Table 1depicts the coverage of documents in two forms of a query. Table 2 shows number of common documents in first five and first ten results respectively.

For the data given in table 1, paired *t*-test have been applied, the calculated value of *t* statistics is 0.385 which is less than tabulated value 1.76 for 14 degree of freedom. Thus the null hypothesis is accepted at 5% significance level, that is, there is no significant difference between the coverage of two results.

Table 1. Number of documents retrieved in two forms of a query

| Query pair no. | $x_i$ | $y_i$ |
|---|---:|---:|
| Q1 | 831,000,000 | 201,000,000 |
| Q2 | 67,100,000 | 372,000,000 |
| Q3 | 134,000,000 | 42,400,000 |
| Q4 | 1,080,000,000 | 2,450,000,000 |
| Q5 | 17,100,000 | 224,000,000 |
| Q6 | 36,800,000 | 371,000,000 |
| Q7 | 575,000,000 | 405,000,000 |
| Q8 | 22,400,000 | 20,500,000 |
| Q9 | 227,000 | 714,000 |
| Q10 | 15,000,000 | 14,600,000 |
| Q11 | 75,600,000 | 75,700,000 |
| Q12 | 19,700,000 | 11,200,000 |
| Q13 | 15,100,000 | 19,600,000 |
| Q14 | 1,400,000 | 8,680,000 |
| Q15 | 1,400,000,000 | 758,000,000 |

Table 2. Number of common documents in first five ($D_5$) and first ten ($D_{10}$) retrieved documents

| Query pair no. | $D_5$ | $D_{10}$ |
|---|---|---|
| Q1 | 3 | 3 |
| Q2 | 2 | 4 |
| Q3 | 4 | 5 |
| Q4 | 2 | 2 |
| Q5 | 3 | 7 |
| Q6 | 3 | 6 |
| Q7 | 4 | 5 |
| Q8 | 4 | 6 |
| Q9 | 3 | 8 |
| Q10 | 2 | 3 |
| Q11 | 3 | 4 |
| Q12 | 2 | 5 |
| Q13 | 4 | 7 |
| Q14 | 4 | 8 |
| Q15 | 4 | 5 |

For the data given in column 2 of table 2, we applied *t*-test for sample mean; the calculated value of *t* statistics is 8.37 which is greater than tabulated value 1.76 for 14 degree of freedom. Thus the null hypothesis is rejected at 5 % significance level, that is, there is significant difference between the sample mean and the population mean. Thus, first five documents in two results are significantly different.

For the data given in column 3 of table 2, we again applied *t*-test for sample mean; the calculated value of *t* statistics is 9.86 which is greater than tabulated value 1.76 for 14 degree of freedom. Thus the null hypothesis is rejected at 5 % significance level, that is, there is significant difference between the sample mean and the population mean. Thus, first ten documents in two results are significantly different.

## 3. CONCLUSIONS

The experiment on Google search results has been performed to check the ability of search engine for responding over a pair of semantically same but different structural queries. In this work, we have tried to check whether common user is getting same results for a query asked in two different ways or not. According to our experiment, there is no significant difference between the coverage of two results, this shows that the search engine provides almost same number of results for a query asked in any form but first five and first ten results of two queries are significantly different. As from the previous researchers, it has been observed that most of the user check the first page, hence it can be concluded that a common user does not get same results for a query when asked in different ways. To get optimum results one should modify one's query in every possible way because every modification provides a chance to get new results. It also signifies the inability of the search engine for providing results based on semantic structure of a sentence which can open a new dimension for researchers in this field.

**Appendix A. List of pairs of Queries**

| | | | |
|---|---|---|---|
| Q.1 | Indian Economy | / | Economy of India |
| Q.2 | Car Accident | / | Accident of car |
| Q.3 | Diabetes Diet | / | Diet for Diabetes |
| Q.4 | Office Management | / | Management in Office |
| Q.5 | Finance Project Report | / | Project Report on finance |
| Q.6 | Kids fun games | / | Fun games for kids |
| Q.7 | Statistics Books | / | Books on Statistics |
| Q.8 | Income tax return filing procedure | / | Procedure for income tax return filing |
| Q.9 | Kumaon Himalayas | / | Himalayas of Kumaon |
| Q.10 | Human behaviour Analysis | / | Analysis of human behaviour |
| Q.11 | Wildlife survey | / | Survey on wildlife |
| Q.12 | Ancient Indian History | / | History of Ancient India |
| Q.13 | Moral Values stories | / | Stories on moral values |
| Q.14 | Financial sector reforms in India | / | Reforms in financial sector in India |
| Q.15 | Health care policy issues | / | Policy issues in health care |